\documentclass[floatfix,twocolumn,showpacs,longbibliography,nofootinbib,superscriptaddress,prb]{revtex4-1}
\usepackage[T1]{fontenc}
\usepackage[utf8]{inputenc}
\usepackage{lmodern}
\usepackage{graphicx}
\usepackage{amsmath}
\usepackage{xcolor}
\usepackage{soul}
\usepackage{float}
\usepackage{bm}
\usepackage{comment}
\renewcommand{\vec}[1]{\mathbf{#1}}
\usepackage{hyperref}
\makeatletter
\newcommand*{\rom}[1]{\expandafter\@slowromancap\romannumeral #1@}
\makeatother

\begin{document}
\renewcommand{\vec}[1]{\mathbf{#1}}
\newcommand{\ii}{\mathrm{i}}

\title{Topological entanglement entropy to identify topological order in quantum skyrmions}

\author{Vipin Vijayan}
\affiliation{Department of Physics, Indian Institute of Technology (Banaras Hindu University), Varanasi 221005, India }
\author{L. Chotorlishvili}
\affiliation{Department of Physics and Medical Engineering, Rzesz\'ow University of Technology, 35-959 Rzesz\'ow, Poland}
\author{A. Ernst}
\affiliation{Max Planck Institute of Microstructure Physics, Weinberg 2, D-06120 Halle, Germany}
\affiliation{Institute for Theoretical Physics, Johannes Kepler University, Altenberger Stra{\ss}e 69, 4040 Linz, Austria}
\author{S.\,S.\,P. Parkin}
\affiliation{Max Planck Institute of Microstructure Physics, Weinberg 2, D-06120 Halle, Germany}

\author{M.\,I. Katsnelson}
\affiliation{Radboud University, Institute for Molecules and Materials, Heyendaalseweg 135, 6525AJ Nijmegen, Netherlands}

\author{Sunil K. Mishra}
\affiliation{Department of Physics, Indian Institute of Technology (Banaras Hindu University)}

\date{\today}
\begin{abstract}
\textcolor{black}{We study the topological entanglement entropy and scalar chirality of a topologically ordered skyrmion formed in a two-dimensional triangular lattice. Scalar chirality remains a smooth function of the magnetic field in both helical and quantum skyrmion phases. In contrast, topological entanglement entropy remains almost constant in the quantum skyrmion phase, whereas it experiences enhanced fluctuations in the helical phase. Therefore, topological entanglement entropy is an effective tool to distinguish between the two phases and pinpoint the quantum phase transition in the system.}

\end{abstract}

\maketitle

\textcolor{black}{Topological phases are robust against weak perturbations(disorder and interactions which preserve the symmetry, on average) due to
their bulk gap.\cite{2021NatRM...7..196W}. This is referred to as topological protection.} They are characterized by distinct values of the
topological invariant. This quantity stays constant under smooth
deformations when the system is in a topological phase. Even when the
actual state of the system varies, the invariant remains the same
within the phase. Smooth deformations are defined as the kind of
changes in the Hamiltonian that do not close the bulk energy gap~\cite{2011RvMP...83.1057Q}.

Among topological materials, the \textcolor{black}{latest and highly promising are skyrmions.} They have peculiar non-collinear spin
textures~\cite{Bogdanovvortex1989_1,Bogdanovvortex1989_2,Bogdanovvortex1994,piette1995multisolitons,rajaraman1982solitons,skyrme1994non}.
Despite the small size, non-trivial topological characteristics
confer them with a remarkable degree of thermal stability even at
room-temperature~\cite{2016NatNa..11..444M,2016NatNa..11..449B,2016NatMa..15..501W,PhysRevB.104.054420,PhysRevLett.127.097201,fret2013_skyr,2021NatCo..12.5079Y}. Due to these properties, skyrmions are considered to be a candidate for
information carriers in next-generation spintronic
devices~\cite{PhysRevB.87.134403,2020JPCM...32n3001Z,2020NatRP...2..492B}. \textcolor{black}{Recently, the name "Skyrmionics" has drawn a lot of attention.}
Simply explained, a skyrmion can be described as a nonlinear localized
mode with a stable shape. 
 Quantum skyrmions have a history of over sixty years
\cite{skyrme1962unified,belavin1975metastable}, whereas the magnetic skyrmions have been studied classically with great success since pioneering works \cite{Bogdanovvortex1989_1}. \textcolor{black}{Tony Skyrme introduced the concept of skyrmions in a series of papers published between 1958-1962. He hypothesized that fermionic nucleons could be produced from bosonic pion fields \cite{2020arXiv200109944A}. He also discussed a non-trivial topological structure distinguished by a parameter called the topological invariant. This idea did not gain widespread recognition at the time. Later on, it was realized that topologically non-trivial magnetic skyrmions exist. These have a characteristic invariant called the topological number, which is the number of times the spin directions wrap around the surface of a sphere \cite{doi:10.1146/annurev-conmatphys-031620-110344}.} The nanoscale of magnetic skyrmions made it important to study quantum
properties too~\cite{2022arXiv220110702B,doi:10.1126/science.1240573,2011NatPh...7..713H}.
So far there are only  a few works which tried to describe magnetic skyrmions quantum
mechanically~\cite{PhysRevX.9.041063,PhysRevB.103.L060404,PhysRevB.107.L100419,PhysRevResearch.4.023111,PhysRevB.94.134415,ochoa2019quantum,PhysRevX.7.041045,
zhou2020solids,PhysRevB.103.224423,PhysRevResearch.4.043113}. \textcolor{black}{A quantum mechanical description of a skyrmion phase is far from complete without identifying a distinguishable topological invariant whose origin is purely from the related wave function}.
The problem is, however, that quantum skyrmion is not
rigorously speaking, strictly topologically protected
in the same way, as the classical skyrmion is. In a
less formal way, one can say that zero-point quantum
fluctuations mix a skyrmionic configuration with
a topologically more trivial one, similar to the decrease
of the sublattice magnetization due to zero-point magnon
fluctuations in an antiferromagnet~\cite{Sotnikov2022}. In this context,
many quantities which may play the role of \textcolor{black}{a topological marker} were explored.

By \textcolor{black}{measuring the} quantum scalar chirality Sotnikov {\it et.al.} could
identify a skyrmion phase in the ground state of a triangular spin
lattice with nearest-neighbor Heisenberg antiferromagnetic coupling
and the Dzyaloshinskii-Moria interaction (DMI)~\cite{PhysRevB.103.L060404}. They
used scalar chirality as an \textcolor{black}{analog} to the classical topological charge
to reason their report. But we see that the scalar chirality fails in
distinguishing the helical phase and a skyrmion with absolute precision.
\textcolor{black}{The reason is that the skyrmion phase is essentially a helical phase,} resulting in a smooth transition between both. Siegl {\it et. al.} have used the winding parameter and the topological index to quantify the topological protection~\cite{PhysRevResearch.4.023111} in the ground state and excited states. The winding
parameter behaves just like the scalar chirality used by Sotnikov {\it et. al.}. The topological index turned out to be zero for the system
that we are working with. This is because the average expectation
value of the X, Y, and Z components of the spin operator turned out to be
the same for all 19 spins in the system. This will become clearer later. \textcolor{black}{Makuta {\it et al.} \cite{2023arXiv230408199M} have shown various quantum skyrmions in a Hubbard model with strong spin-orbit coupling. Haller {\it et al.} shows the magnetic texture and related properties of quantum skyrmions formed in various lattices using MPS and DMRG \cite{PhysRevResearch.4.043113}. Joshi {\it et al.} \cite{2023arXiv230409504J} use a novel neural network quantum state to approximate the ground state of a quantum skyrmion in a model similar to that of Siegl {\it et al.}}. 

\textcolor{black}{Skyrmions are topological entities, and the entanglement entropy in the ground state is expected to reflect this as some sort of universal feature \cite{PhysRevLett.96.110404}.
There are no accounts for the quantum skyrmion in this perspective in
classifying the helical and the skyrmion phases. We are addressing this
issue in our work.}

Topological features of a system can impact the entanglement properties 
\cite{PhysRevB.107.115126,PhysRevB.106.224418}.
The ground state associated with a topological phase is known to show
robust entanglement properties. In the light of this knowledge, we studied topological entanglement entropy (TEE). 
TEE has been
calculated in toric code systems
\cite{2003AnPhy.303....2K,2005PhLA..337...22H,2007PhRvB..76r4442C,PhysRevLett.96.110405}
and in a honeycomb lattice
\cite{2006AnPhy.321....2K,2010PhRvL.105h0501Y}.
In this work, we will focus on the effects of topology on the
entanglement entropy and related quantities of a skyrmion system. We
will look at the behavior of these quantities around the critical
points of the phase transition and within the skyrmion phase.

\textcolor{black}{The quantum states associated with the ferromagnetic phase are devoid of entanglement, }while quantum phases with noncollinear spin order (helical or quantum skyrmion phase) have complex entanglement properties. It is easy to differentiate between product and entangled states, but it is challenging to determine the quantum phase transition between helical and quantum skyrmion phases. However, our research shows that TEE precisely indicates this transition. Adjusting the external magnetic field can drive the system from the helical to the quantum skyrmion phase. The TEE behaves differently in each quantum phase. In the helical phase, TEE shows enhanced fluctuations and a smooth plateau in the quantum skyrmion phase. These differences in TEE enable us to identify both quantum phases and the border between them.  
 
The main results that We report are TEE along with the scalar chirality can be used
to accurately distinguish various topological phases in a helical
magnetic system. In a skyrmion system we show that increasing
antiferromagnetic next-nearest neighbor interaction results in
topological protection against a large applied magnetic field. \textcolor{black}{Increasing the parameter $J_2$ results in decreasing the correlation length of the system, which in turn helps to improve the invariance of TEE in the skyrmion phase.}

\begin{figure}[ht]
  \includegraphics[width=0.1\textwidth]{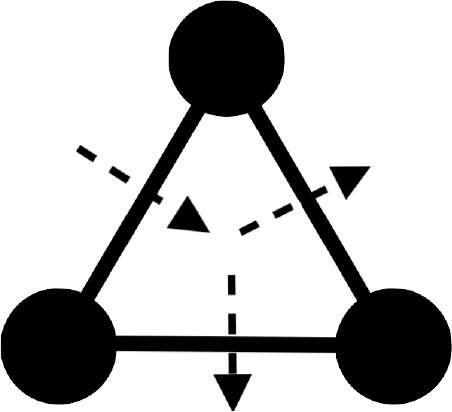}
  \caption{Three bond types(solid lines connecting the dots) and
    direction of corresponding DMI vectors(dashed arrows).}
\label{fig:lattice}
\end{figure}

{\it Spin Model:}- We consider a configuration of spins formed in a triangular lattice
with periodic boundary conditions (PBC) (see
Fig.~\ref{fig:lattice}). This configuration possesses a six-fold
rotational symmetry and translation symmetry.

A formation of non-collinear magnetic textures like skyrmions are
mainly governed by the competing interaction within some spin
systems. The competing interactions are either nearest-neighbor
ferromagnetic and next-nearest neighbor antiferromagnetic
interactions or nearest-neighbor ferromagnetic and the 
DMI~\cite{belavin1975metastable,barton2020magnetic,schroers1995bogomol,seki2012observation,wilson2014chiral,
  schutte2014magnon,white2014electric,derras2018quantum,haldar2018first,leonov2015multiply,
  psaroudaki2017quantum,van2013magnetic,rohart2016path,samoilenka2017gauged,battye2013isospinning,
  jennings2014broken,tsesses2018optical}. The DMI has a key role in
creating and stabilizing skyrmions. \textcolor{black}{We consider that
the Hamiltonian of the spin-$\frac{1}{2}$ system has the form:}
\begin{eqnarray} 
\label{DM-skyrmion}
 \hat H=B\sum\limits_i \hat S_i^z+J_1\sum\limits_{\langle i,j\rangle}\hat{\vec{S}}_i \hat{\vec{S}}_j+&
  J_2\sum\limits_{\langle\langle i,j\rangle\rangle}\hat{\vec{S}}_i\hat{\vec{S}}_j\nonumber\\
 +\sum\limits_{ij}\textbf{D}_{i,j}\left[\hat{\vec{S}}_i\times\hat{\vec{S}}_j\right], 
\end{eqnarray}
Here we consider both $J_1$(ferromagnetic nearest-neighbour exchange)
and $J_2$ (antiferromagnetic next-nearest neighbor exchange)
interactions along with the DMI. \textcolor{black}{Single angle brackets denote
summation over unique pairs of  nearest neighbors and double angle brackets denote
summation over unique next-nearest neighboring pairs.} Here $B$ is an external
magnetic field and the DM vector \textcolor{black}{$\bm{D_{i,j}}$ is aligned perpendicular
to the bond between lattice sites $i$ and $j$ (see the Fig.~\ref{fig:lattice}). \it{i.e,} $\bm D_{i,j} = D \hat e_z\times(\bm r_j - \bm r_i)$ where $D = |\bm{D_{i,j}}|$.} We note that in the helical
multiferroic insulators, the constant $D$ can be expressed in terms of
the magneto-electric coupling $g_{ME}$ with an external electric
field $E$, {\it i.e.}, $D=g_{ME}E$. Thus, one can control the strength
of the DMI term externally~\cite{PhysRevLett.125.227201,wang2020optical}.
By adjusting the intensity of a high-frequency laser, the DMI interaction can be enhanced. This method involves dynamic control of the intrinsic magnetic interactions through the electronic tunneling processes in specific fields. Ultimately, this method leads to the effective rescaling of DMI and exchange constants in the desired way \cite{PhysRevLett.118.157201,PhysRevLett.115.075301}. 
The above model with the Hamiltonian as in Eq.~(\ref{DM-skyrmion}) \textcolor{black}{yields peculiar phases
in the parameter space}. This setting creates the
same environment around each spin. Because of this, the physical
observables measured on a single spin will give the same outcome for
all 19 spins. The DMI term compels the spins to lie in the plane
while making a rotation in the local magnetic texture, while the
$J_1$, $J_2$ terms encourage the parallel orientation of spins. When a
magnetic field is applied, the system tends to form non-collinear
magnetic textures. \textcolor{black}{We notice that the lower energy levels of the system show different degeneracies through various phases. It is shown in supplementary materials \cite{[{See }][{at [URL will be added by the publisher] for detailed discussion about degeneracy, zoomed in plots of {TEE} where boundaries are better visible, the dependance of correlation length on the parameter {$J_2$} and the fidelity between states in different phases.}]supplementary} Fig.~1 how degeneracy $\Gamma$ varies in different phases. The expectation values that we calculate are averaged over these degenerate states whenever it is required.  For interested readers, we also included the results in the multiplet of degenerate states in the supplementary material.} We explored the following quantities on this model for the
construction of this paper.

{\it Methods:-} The formation of a quantum non-trivial magnetic structure can be
investigated using spin correlation functions. In particular, we explore
the Fourier transform of a longitudinal spin correlation function $G_{\|}(\textbf{q})=\dfrac{1}{N}\sum\limits_{ij}G_{\|}(\textbf{r}_{ij})\exp(-i\textbf{q}\textbf{r}_{ij})$,
where $N$ is the number of spins and the correlation function reads:
\begin{eqnarray}
\label{correlation functions}
G_{\|}(\textbf{r}_{ij})=\langle n\vert\hat S_i^z\hat S_j^z\vert n\rangle.
\end{eqnarray}
\textcolor{black}{Here, $n$ denotes the ground state vector. We expect to see intensity peaks or the Braggs spots in  $G_{\|}(\textbf{q}$) for nonzero
\textbf{q} values as a result of the existence of non-trivial magnetic
textures~\cite{stepanov2019heisenberg}.}

Next we look at the quantum scalar chirality($C_\psi$). $C_\psi$ is considered as a topological invariant to
distinguish between a quantum skyrmion (topological) phase and other phases of the system \cite{PhysRevB.103.L060404}. \textcolor{black}{For a spin-$\frac{1}{2}$ system},
\begin{eqnarray} 
\label{DM-skyrmion_chirality}
&& \textbf{C}_\Psi =\frac{L}{\pi}\left\langle \hat{\vec{S}}_1\left[ \hat{\vec{S}}_2\times\hat{\vec{S}}_3\right] \right\rangle.
\end{eqnarray}
Here $L$ is the number of non-overlapping elementary triangular
patches covering the lattice and three adjacent spins form a
patch. The scalar chirality for any of the three adjacent spin
combinations is the same. This is because of the translational and
rotational symmetries of the lattice. This quantity also indicates the
stability of the corresponding skyrmion phase against
perturbations. 

Our main goal is to look for a correlation between topological invariance and entropy of the system.
 The entanglement entropy~(EE) of a system is the measure of entanglement degree between two of its
 composite subsystems.  
 Here we will divide our system into subsystems $Q$ and its
 complement, $\Tilde{Q}$. The composite Hilbert space $H_S$ of the
 system is a tensor product of the local Hilbert spaces $H_Q$ and
 $H_{\Tilde{Q}}$. Then the entanglement entropy is calculated by
 considering the reduced density matrix of the subsystem $Q$, $\rho_Q$, and
 calculating the von Neumann entropy. 
 
 \begin{equation}
     \mathcal{S}_Q=-tr (\rho_Q \log (\rho_Q))
 \end{equation}

 A nonzero value of $\mathcal{S}_Q$ tells us that the two subsystems are entangled. There are quantities which are derived from entanglement entropy that encapsulates the essence of topological nature of the system, like Topological entanglement entropy~(TEE).

\begin{figure}[!ht]
  \centering
    \includegraphics[width=.417\linewidth]{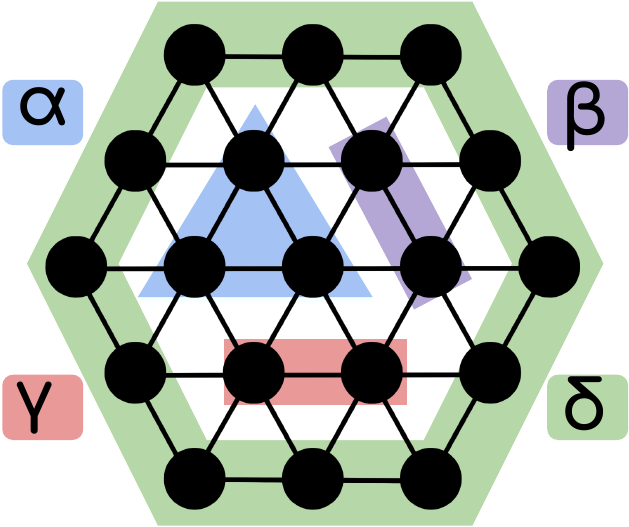}
  \caption{Partitioning of the system for calculating the
    TEE. Different colors identify corresponding subsystems
    $\alpha, \, \beta, \, \gamma$ and $\delta$.}
  \label{Fig:partitioning}
\end{figure}

TEE, ($\mathcal{S}_{topo}$), is
defined as a universal characterization of the many-particle quantum
entanglement in the ground state of a topologically ordered
two-dimensional medium with a mass gap~\cite{PhysRevLett.96.110405,
 PhysRevLett.96.110404}. The bipartite entanglement entropy or the
entanglement entropy(EE) measures the extent to which one partition of
a system is entangled with the next one of the two partitions. It has
been found that the EE obeys an 'area law' {\it i.e,} the entanglement
\textcolor{black}{entropy is proportional to the area partition boundary \cite{PhysRevLett.71.666}}. Kitaev and Preskill \cite{PhysRevLett.96.110404} and
Levin and Wen \cite{PhysRevLett.96.110405} in their independent works
showed that apart from the area-dependent term there is a part of the
entanglement entropy \textcolor{black}{which is universal and characterize a global feature of entanglement in the ground state.} This term is coined as the topological entanglement
entropy~(TEE). \textcolor{black}{This brief review on TEE is adapted from Kitaev and Preskill\cite{PhysRevLett.96.110404}.} TEE is defined as follows. We consider three partitions
of the plane of the system $\alpha,\beta$, and $\gamma$. The remaining
portion of the system, on which $\alpha,\beta$ and $\gamma$ are
embedded is labelled as $\delta$
(see Fig.~\ref{Fig:partitioning}. $\mathcal{S}_\alpha$ is the
entanglement entropy of the reduced density matrix $\rho_\alpha$ of
the subsystem $\alpha$. $\mathcal{S}_{\alpha \cup \beta}$ is the entropy
of the subsystem $\alpha \cup \beta$ with a reduced density matrix
$\rho_{\alpha \cup \beta}$ etc. Here $\cup$ denotes the set operation
union. Then $\mathcal{S}_{topo}$ is defined using the partitions
$\alpha,\beta$ and $\gamma$ as

\begin{equation}
    \mathcal{S}_{topo}=\mathcal{S}_\alpha + \mathcal{S}_\beta + \mathcal{S}_\gamma - \mathcal{S}_{\alpha \cup \beta} - \mathcal{S}_{\beta \cup \gamma} - \mathcal{S}_{\alpha \cup \gamma} + \mathcal{S}_{\alpha\cup \beta\cup \gamma}.
\end{equation}

For a system with boundary of length $B$, large compared to the correlation length, $\mathcal{S}_{topo}$ is defined to be invariant
under smooth deformations of the Hamiltonian and will only change when
a quantum critical point is encountered. \textcolor{black}{The value of this invariant also depends on the topology of $\alpha$, $\beta$ and $\gamma$. $\mathcal{S}_{topo}$ is not always universal in nature.  In a recent work by Kim {\it et.al} \cite{PhysRevLett.131.166601} shows that even though $\mathcal{S}_{topo}$ can differ in two states related by constant-depth circuits that are in the same phase, it is still valid as a universal lower bound.}

{\it Results and discussion:-} \textcolor{black}{Now let us look at the results of} the various quantities in the chiral system described previously. \textcolor{black}{We propose a systematic approach to characterize
the system into various phases like the quantum skyrmion phase.} First, we discuss the
Fourier transform of the longitudinal spin correlation function.

\begin{figure*}[!ht]
  \centering
  \begin{tabular}{c|c}
$J_2=0.2|J_1|$ & $J_2=0.8|J_1|$
  \\ \includegraphics[width=.36\linewidth]{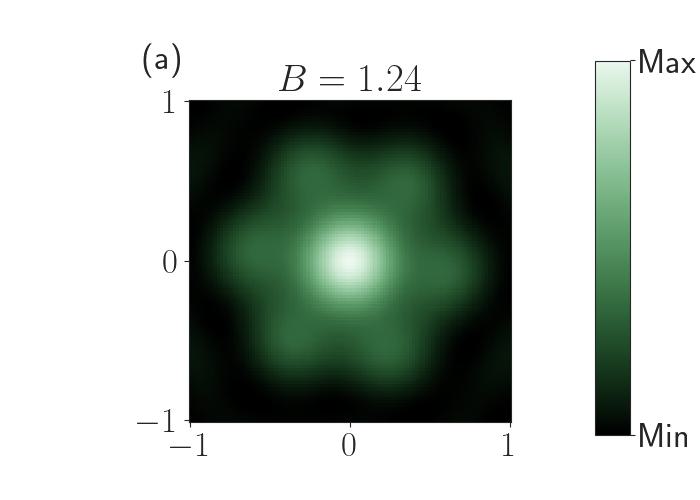} & \includegraphics[width=.36\linewidth]{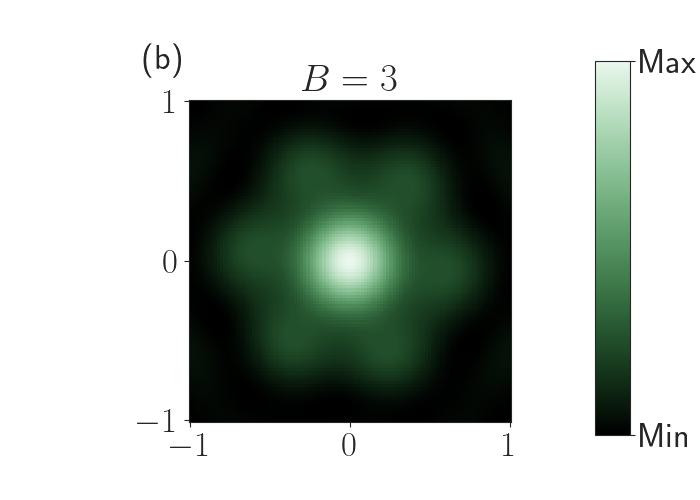} 
    
    \\\includegraphics[width=.32\linewidth]{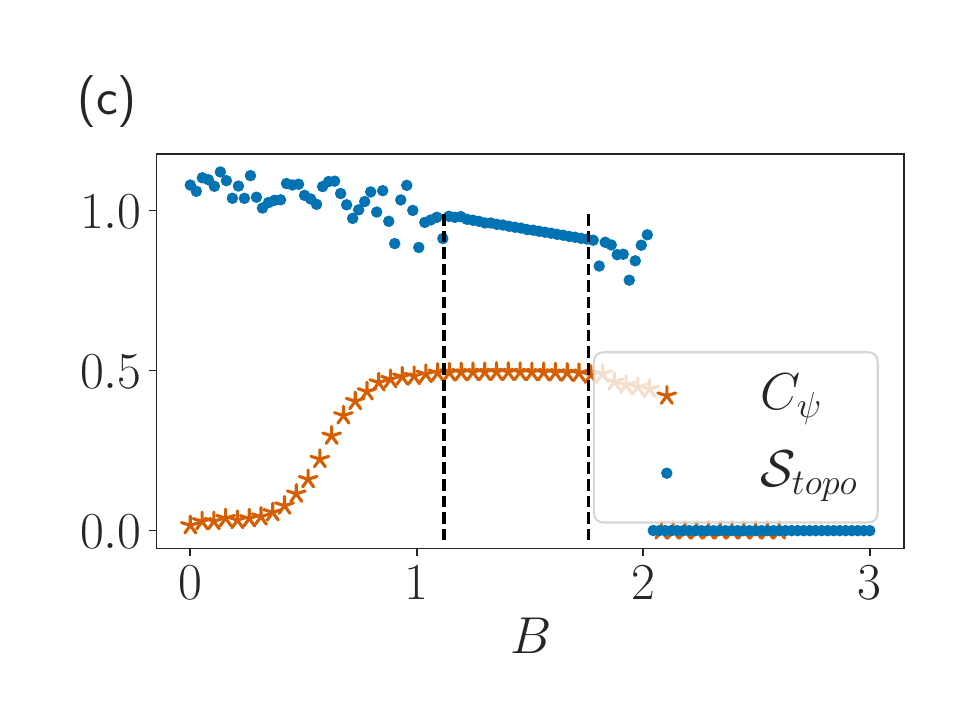} & \includegraphics[width=.32\linewidth]{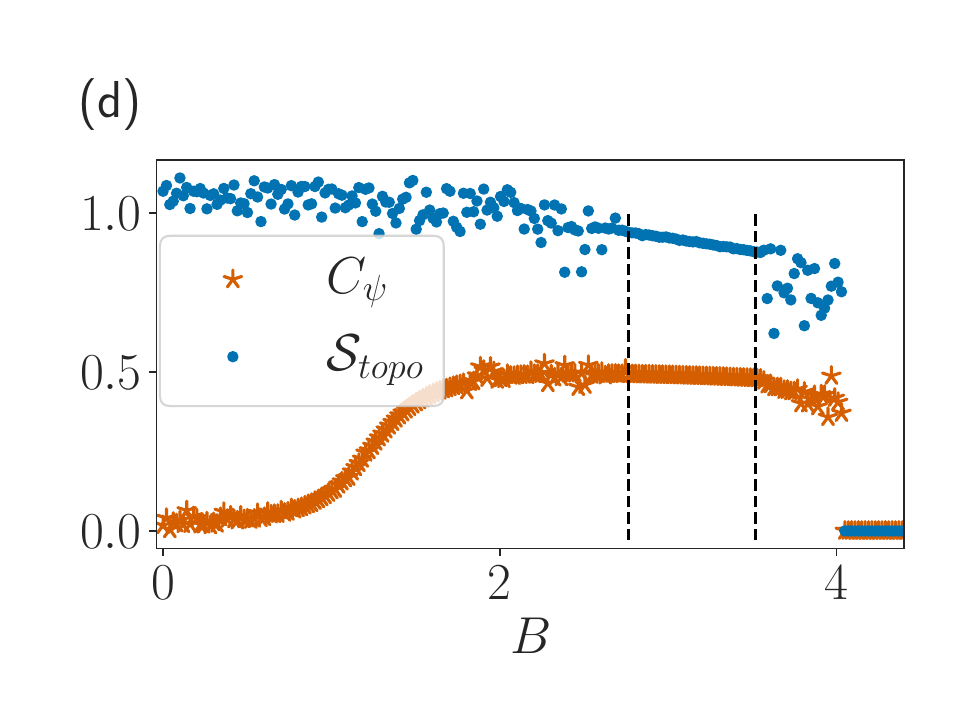}

    \\\includegraphics[width=.32\linewidth]{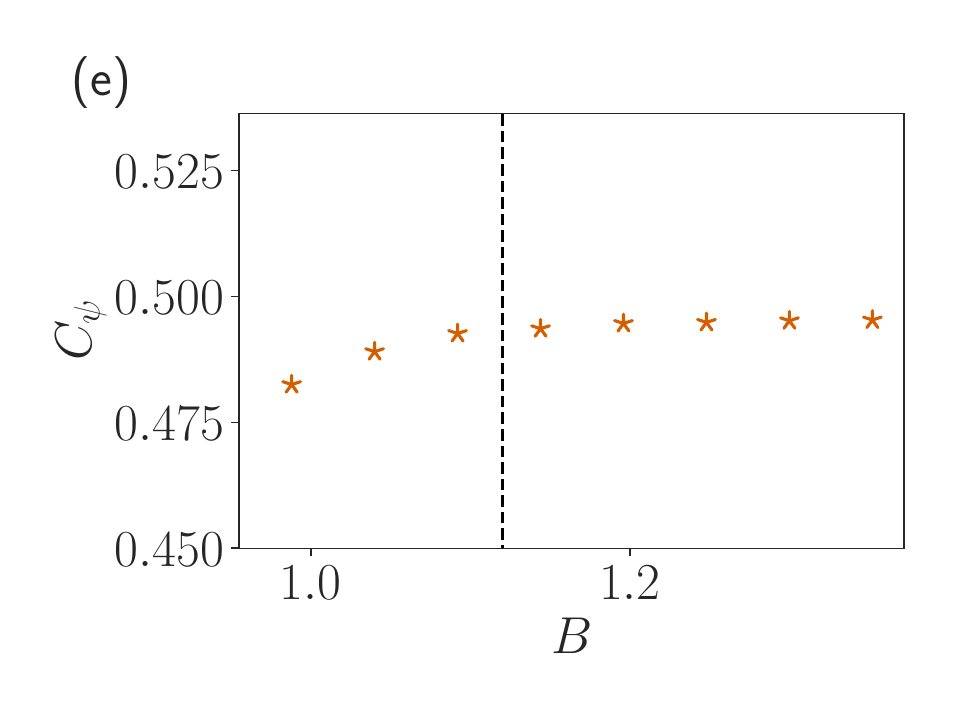} &
  \includegraphics[width=.32\linewidth]{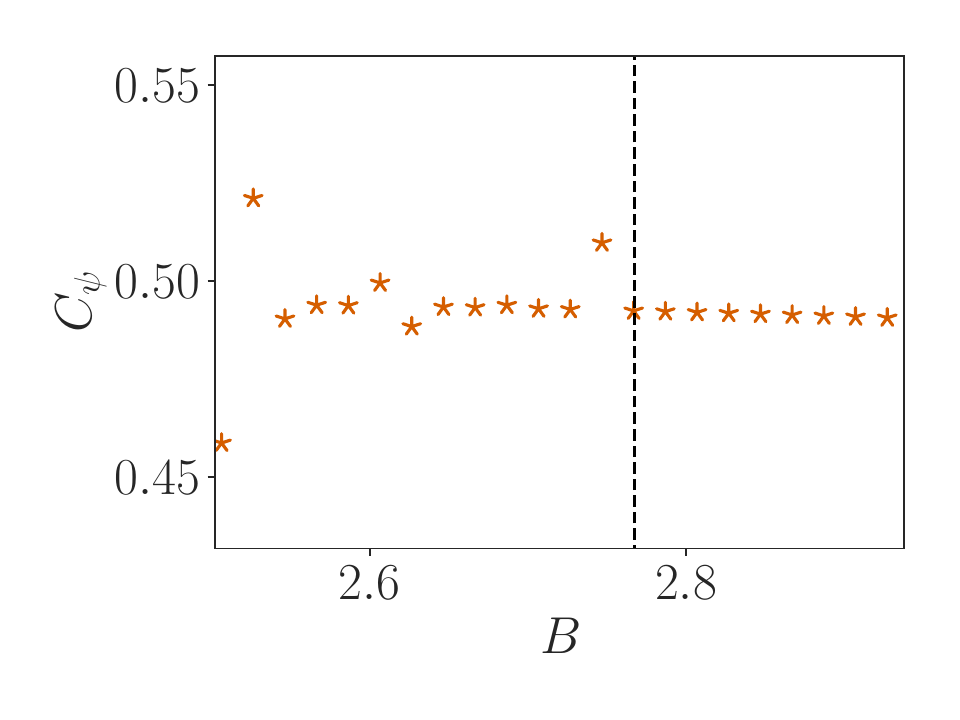} 

  \end{tabular}
  \caption{(a) Fourier transform of the longitudinal structure factor
    calculated for $J_2=0.2|J_1|, \; B=1.24|J_1|$. Bright peaks show
    non-collinear magnetic textures. (b) Fourier transform of the
    longitudinal structure factor calculated for $J_2=0.8|J_1|, \; B=3.0|J_1|$
    showing non-collinear magnetic texture.  (c) TEE and
    $\mathbf{C_\psi}$ vs an applied field for $J_2=0.2|J_1|$. The linear section
    of the TEE aligns perfectly with a plateau of $C_\psi$. The vertical
    lines distinguish between a helical and a skyrmionic phase
    precisely. (d) TEE and
    $\mathbf{C_\psi}$ vs an applied field for $J_2=0.8|J_1|$. The linear section
    of the TEE aligns perfectly with a plateau of $C_\psi$. The vertical
    lines distinguish between a helical and a skyrmionic phase
    precisely. (e) Chirality for $J_2=0.2|J_1|$ zoomed: it is varying slowly before
    the TEE hits the linear part. It becomes constant after crossing
    the boundary set by the TEE(marked by dashed vertical line). (f) Chirality for $J_2=0.8|J_1|$ zoomed, the dashed vertical line is where the TEE becomes linear: it shows
    scattered behavior before the boundary. After the TEE hits the
    linear part, the chirality becomes constant. For all the graphs,
    $D=2|J_1|, \; J_1=-1$. 
    }
\label{Fig:TEE}
\end{figure*}

\begin{figure}[!ht]
  \centering
  \begin{tabular}{cc}
  \includegraphics[width=.48\linewidth]{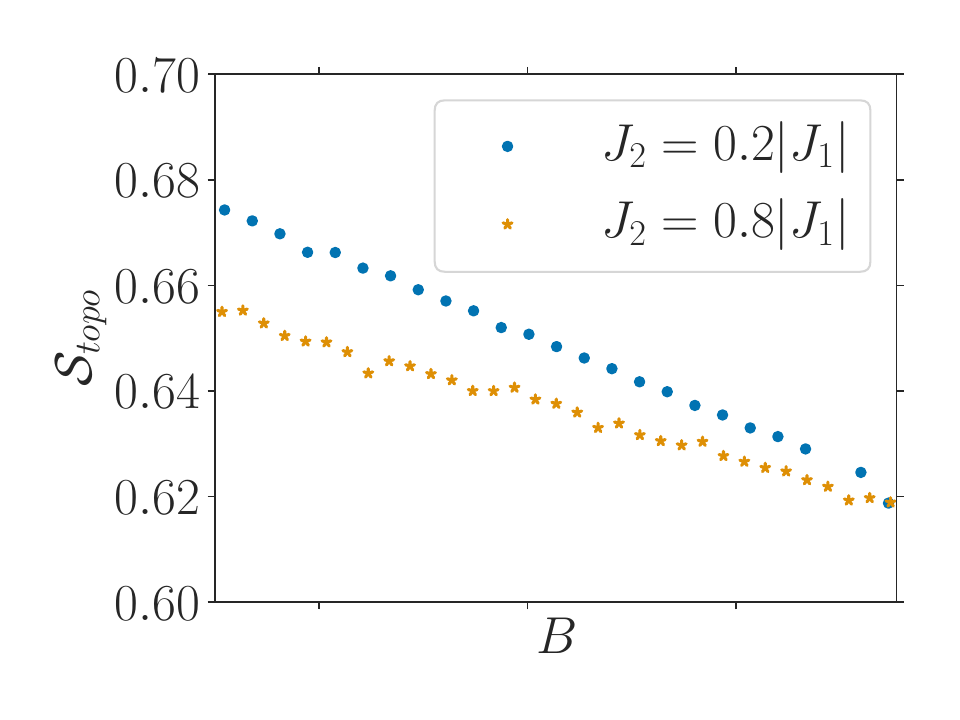} &
  \includegraphics[width=.528\linewidth]{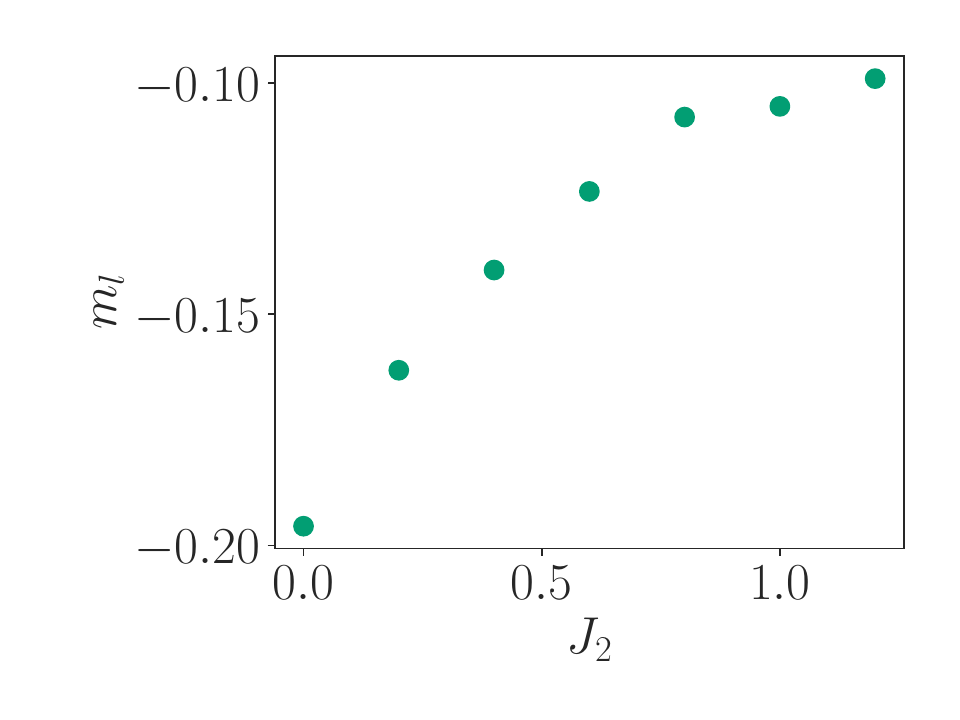} 
   \\ (a) & (b) 
  \end{tabular}
  \caption{(a) The linear part of TEE plotted for two different
      $J_2$s. This shows that for larger $J_2$ the slope is
    less. Note that in this plot the axes do not signify absolute
    values of the quantities, the purpose is to compare the slopes
    only. (b) The slope of the linear part of TEE, $m_l$ vs the parameter
      $J_2$. $D=2|J_1|, \; J_1=-1$ are used for all of the graphs.} 
  \label{Fig:TEE_slope}
\end{figure}

In Figs.~\ref{Fig:TEE}(a) and \ref{Fig:TEE}(b) we look for the \textcolor{black}{Braggs}
peaks at nonzero \textbf{q} values in the Fourier transform of the
longitudinal correlation function for two different sets of
parameters, as an indicator of non-collinear spin textures. The
presence of peaks confirms the helical nature of the system. Note that
we are not able to distinguish a skyrmion from a helical state from
the correlation function. \textcolor{black}{In order to confirm the existence of a
skyrmion phase we use Eq.~(\ref{DM-skyrmion_chirality}) and}
compare it with the TEE to determine the boundary between
various phases. The quantum scalar chirality, which is the quantum
variant of classical chirality, shows three phases of the ground
state. We can clearly distinguish between a chiral phase and a trivial
ferromagnetic phase (see Fig.~\ref{Fig:TEE}: (c,d) orange curve is
the scalar chirality and the zero corresponds to the trivial phase). \textcolor{black}{Here the transition from the chiral to the
skyrmion phase is slow(smooth and not distinguishable) and we can not clearly pinpoint a boundary
between them.}

Following the concepts from earlier established
work(Ref.~\onlinecite{PhysRevB.103.L060404}), the plateau of $C_\psi$
can be associated with a skyrmion phase. However, it is difficult to
identify the region of $C_\psi$ where it exactly becomes flat or
parallel to the x-axis(see Fig.~\ref{Fig:TEE} (c,d)
orange curve). For instance in Fig.~\ref{Fig:TEE} (c) at the applied
field $B\sim 1.0|J_1|$ the chirality seems to hit the plateau. But in
reality, the value is still changing slowly. Although, the chirality
demonstrates the robustness of a given skyrmion phase, it fails to
show a clear boundary between different phases. Therefore, only
$C_\psi$ can not be used to determine the topological phases in a
skyrmion system since it is difficult to pinpoint where the helical
phase ends and where the skyrmionic phase begins. At the same time,
the TEE shown along with $C_\psi$ in Fig.~\ref{Fig:TEE}(c) \textcolor{black}{is
characterized by a continuous and fluctuation-free curve, }which is associated with a topologically protected phase of the skyrmion. Thus, it is much easier to identify the three phases using the TEE: (i) the TEE is zero
for the trivial phase; (ii) it is scattered for a helical phase; (iii)
it is continues for a skyrmionic phase. In Fig.~\ref{Fig:TEE}(d), we
have the same quantities plotted for a larger $J_2$(0.8$|J_1|$)
value. Here we can see that, for larger next-nearest antiferromagnetic
interaction, the topological phase remains robust against a stronger
applied magnetic field. Here also, the TEE helps to identify a
transition from a helical to a skyrmionic phase. \textcolor{black}{Figures showing zoomed view to the phase boundaries of TEE is provided in the supplementary file.} The
slow variation of the scalar chirality before crossing the critical
boundary established by TEE can be recognized in Fig.~\ref{Fig:TEE}(e), and
the constant plateau of chirality can be seen after the critical point
in Fig.~\ref{Fig:TEE}(f).
 
For a two-dimensional system having a topologically protected phase,
the TEE would show a constant plateau around the corresponding applied
field if the system boundary is large enough \textcolor{black}{compared to the correlation length}. Due to the finite size limit,
this is not realizable in this system. Because of computational
limitations making a scaling analysis is currently
difficult. We attempted to use state-of-the-art approximation techniques, including PEPS, DMRG, variational quantum Monte Carlo, and neural network quantum states, and report the results in Supplementary Section V. Our goal was to identify techniques suitable for larger system sizes. However, none of these methods converged to the ground state for the 19-spin case.

Nevertheless, if we compare the slopes, \textcolor{black}{$m_l=\frac{\Delta(\mathcal{S}_{topo})}{\Delta(B)}$,} of the linear part
of the TEE, For two different values of $J_2$, we can see that the TEE has a smaller slope for the larger $J_2$ value than for the smaller $J_2$ value. This is show in Fig~\ref{Fig:TEE_slope}(a). \textcolor{black}{the linear parts of the TEE makes angles
$\sim -0.2$ radians and $\sim -0.1$ radians respectively with respect to the positive
x axis)}. This comparison is demonstrated in
Fig.~\ref{Fig:TEE_slope}(a). It is clear that larger $J_2$ results in
a smaller slope. In Fig.~\ref{Fig:TEE_slope}(b), we have plotted slope of the linear
part of TEE vs the parameter $J_2$. We see that the slope of the
linear part of the TEE moves closer to zero as we increase $J_2$. This
tells us that the linear part is approaching a constant plateau as we
increase $J_2$, {\it i.e}, the system is more topologically protected
for larger $J_2$ values. This can be due to the increase of $J_2$
inducing a bulk effect, just as if the system size is increased. \textcolor{black}{In supplementary materials Fig. 7 and the discussion therein we show that the correlation length of the system decreases with increasing $J_2$. One of the conditions considered by Kitaev and Preskill while defining $\mathcal{S}_{topo}$ was large boundary compared to correlation length. Hence increasing $J_2$ helps to better impose this condition too. After a certain limit, the increase of $J_2$ destroys the topological
protection altogether. So care needs to be given while choosing $J_2$.}
  
{\it Conclusion:-} Compared to the ferromagnetic state, quantum states with noncolinear magnetic order are not product states, and they are therefore characterized by nontrivial entanglement properties. Besides, quantum states with noncolinear magnetic order can be diverse. Paradigmatic examples are helical and quantum skyrmion phases. The helical phase is not topologically protected, whereas the quantum skyrmion phase is topologically protected. What is important, in the quantum case, as opposed to the classical case, it is rather demanding to distinguish between these two phases. The reason is that the standard argument about classical magnetic textures does not apply to the quantum case. Consequently, one needs to find other criteria valid to the quantum case. Identifying the quantum phase transition between two helical and quantum skyrmion phases is an even more complicated problem. In this work, we explored one of the crucial properties of a quantum many-body system, entanglement entropy, in the context of the topological protection of a quantum skyrmion system. We used entropic measures of entanglement to classify the skyrmion phase uniquely. We studied two quantities: quantum scalar chirality and topological entanglement entropy. These two quantities we used to identify topological ordering in quantum many-body systems. We note that scalar chirality was already exploited for quantum skyrmions. However, topological entanglement entropy is never discussed in this context before. Though the scalar chirality succeeds in comparing the robustness of two skyrmion phases, it fails to establish a clear boundary between the helical and the skyrmionic phases. In the present work, we discovered that unlike the scalar chirality and the topological index, the topological entanglement entropy precisely 
distinguishes between helical and quantum skyrmion phases and infers 
quantum phase transition between those two phases. Adjusting the external magnetic field can drive the system from the helical to the quantum skyrmion phase. We found that topological entanglement entropy behaves differently in each quantum phase. In the helical phase, it shows enhanced fluctuations and a smooth plateau in the quantum skyrmion phase. These differences enable us to identify both quantum phases and the border between them. Thus topological entanglement entropy can be considered as inherently a topologically exclusive entropic measure of entanglement.
 Finally, based on the topological entanglement entropy, we gave a numerical proof for the observation that the quantum skyrmion model we discussed shows more topological protection when we increase the next nearest exchange coupling $J_2$ within the skyrmionic phase.
{\it Acknowledgments:-} We acknowledge National Supercomputing Mission (NSM) for providing
computing resources of ‘PARAM Shivay’ at Indian Institute of
Technology (BHU), Varanasi, which is implemented by C-DAC and
supported by the Ministry of Electronics and Information Technology
(MeitY) and Department of Science and Technology (DST), Government of
India. A.E. acknowledges funding by Fonds zur Förderung der
Wissenschaftlichen Forschung (FWF) Grant No. I 5384. This project has
received funding from the European Union’s Horizon 2020 research and
innovation programme under Grant Agreement No. 766566 (ASPIN) and No. 854843 (FASTCORR) and from
the Deutsche Forschungsgemeinschaft (DFG, German Research Foundation) - project No. 403505322, Priority Programme (SPP) 2137. 
SKM acknowledges Science and Engineering Research Board, Department of Science and Technology, India for support under Core Research Grant CRG/2021/007095.

\bibliography{2nems}

\end{document}


\renewcommand{\vec}[1]{\mathbf{#1}}
\newcommand{\ii}{\mathrm{i}}

\title{Supplementary to `Topological entanglement entropy to identify topological order in quantum skyrmions'}
\maketitle

\section{Detailed Discussion on degeneracy}

\begin{figure}[H]
    \centering    \includegraphics[width=0.9\columnwidth]{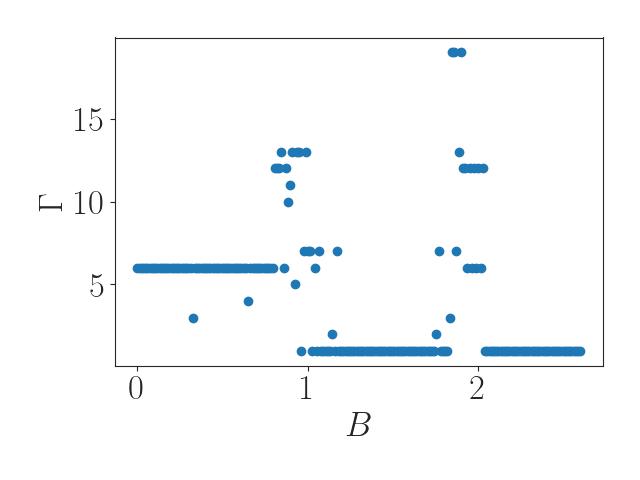}
    \caption{The degeneracy of the ground state, $\Gamma$ is plotted against applied field. We see that at low $B$ values, $GS$ has a degeneracy of $\Gamma=6$. At very high $B$ values, the degeneracy breaks and  $\Gamma=1$. Here for instance we considered $J_1=-1.0$, $J_2=0.2|J_1|$, $D=2.0|J_1|.$}
    \label{fig:degeneracy}
\end{figure}

In Fig.~\ref{fig:degeneracy} we plot degeneracy $\Gamma$ vs magnetic field $B$ for $J_2=0.2|J_1|$, $D=2.0|J_1|$. The plot shows that the helical phase mostly has a six-fold degeneracy. When the system reaches the skyrmion phase from the helical phase, the degeneracy is lifted completely (reduced to a nondegenerate state). It remains mostly nondegenerate in the skyrmion phase. We see that near the critical points of phase transition, the plot shows exotic degenerate levels. The quantities reported in our main text are averaged over these degeneracies.

\begin{figure}[H]
    \centering
    \includegraphics[width=0.9\columnwidth]{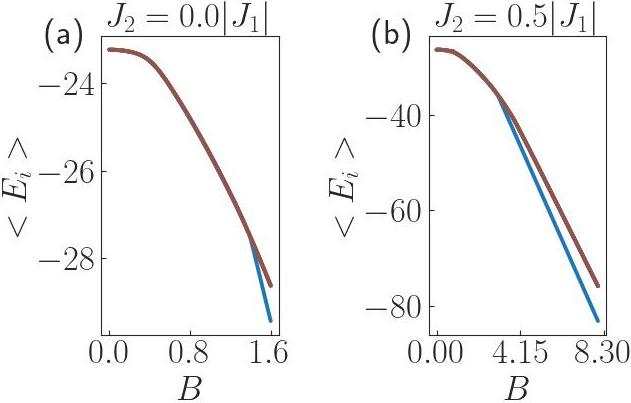}
    \caption{First six eigen energies plotted against the parameter $B$ for (a) $J_2=0.0|J_1|$, $D=2.0|J_1|$ and (b) $J_2=0.5|J_1|$, $D=2.0|J_1|$. We see that they are degenerate up to a certain limit in $B$. In both cases, the degeneracy breaks after a limit when $B$ is increased. }
    \label{fig:energy_degeneracy}
\end{figure}

Fig.~\ref{fig:energy_degeneracy} also indicates the degenerate states explicitly plotting the energy levels vs $B$. Two such instances are shown. With larger $J_2$, we see that the system requires large deformation to break the degeneracy.

\begin{figure}
    \centering
    \begin{tabular}{cc}
        \includegraphics[width=.25\columnwidth]{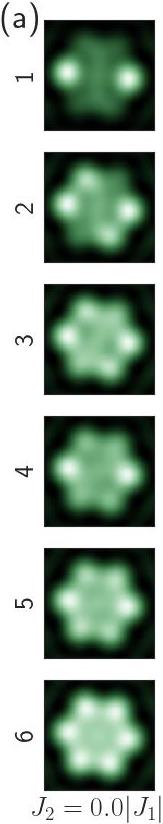} & \includegraphics[width=.25\columnwidth]{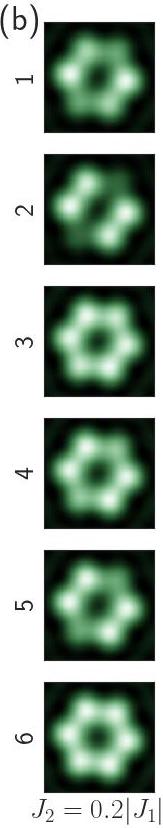} 
    \end{tabular}
    \caption{$G_{\|}(\textbf{q})$ is calculated for each of the 6 degenerate ground state; (a) $J_2=0.0|J_1|$, $B=0.4|J_1|$, $D=2.0|J_1|$ and (b) $J_2=0.2|J_1|$, $B=0.4|J_1|$, $D=2.0|J_1|$. Non-collinearity of spins exists in all these states indicated by bright spots. Both (a) and (b) are instances where the system shows a six-fold degeneracy and is in the helical phase.}
    \label{fig:sf_degenerate}
\end{figure}

$G_{\|}(\textbf{q})$ is plotted for degenerate helical state in Fig.~\ref{fig:sf_degenerate}. We see that each of the degenerate states contributes differently to the noncolinear spin texture. Since the skyrmion and ferromagnetic phases are nondegenerate, we do not require such an illustration for those cases.

\begin{figure}
    \centering
    \includegraphics[width=0.8\columnwidth]{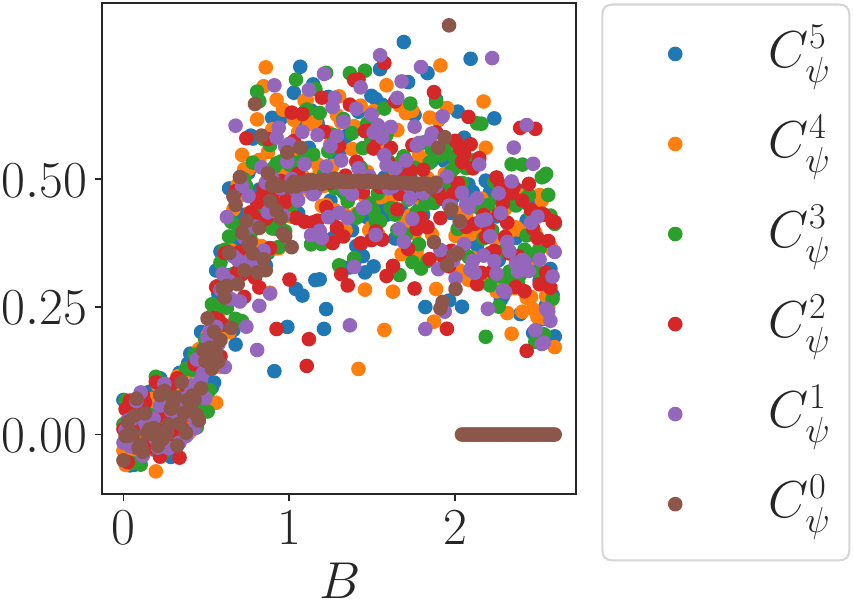}
    \caption{$C_\psi$ first six energy eigen vectors vs $B$ for $J_2=0.2|J_1|$, $D=2.0|J_1|$. }
    \label{fig:chirality_degeneracy}
\end{figure}

We calculated the $C_\psi$ for the first six energy eigenvectors and plotted them together in Fig.~\ref{fig:chirality_degeneracy}. The averaged chirality we report in the main text shows no large fluctuation. In fact, those curves are fairly smooth. But here when we look at individual $C_\psi$ for individual states there appear to be significant fluctuations. Averaging helped to eliminate the fluctuations.

\begin{figure}[H]
    \centering
    \includegraphics[width=0.8\columnwidth]{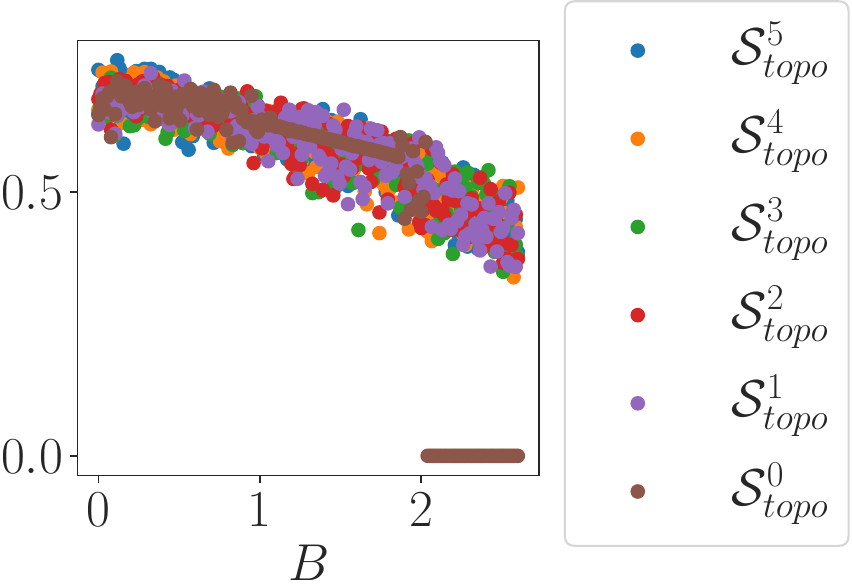}
    \caption{$\mathcal{S}_{topo}$ first six energy eigen vectors plotted against $B$ for $J_2=0.2|J_1|$, $D=2.0|J_1|$. }
    \label{fig:TEE_degeneracy}
\end{figure}

But in the case of $\mathcal{S}_{topo}$, the averaging did not help in removing the fluctuations, as we can see from 
Fig.~\ref{fig:TEE_degeneracy}. From this graph, we suspect that the fluctuation, in fact, is not due to artifact of numerical methods or due to finite size effects.

\section{$\bf{\mathcal{S}_{topo}}$ boundary}

\begin{figure}[H]
    \centering
    \begin{tabular}{cc}
        \includegraphics[width=.48\linewidth]{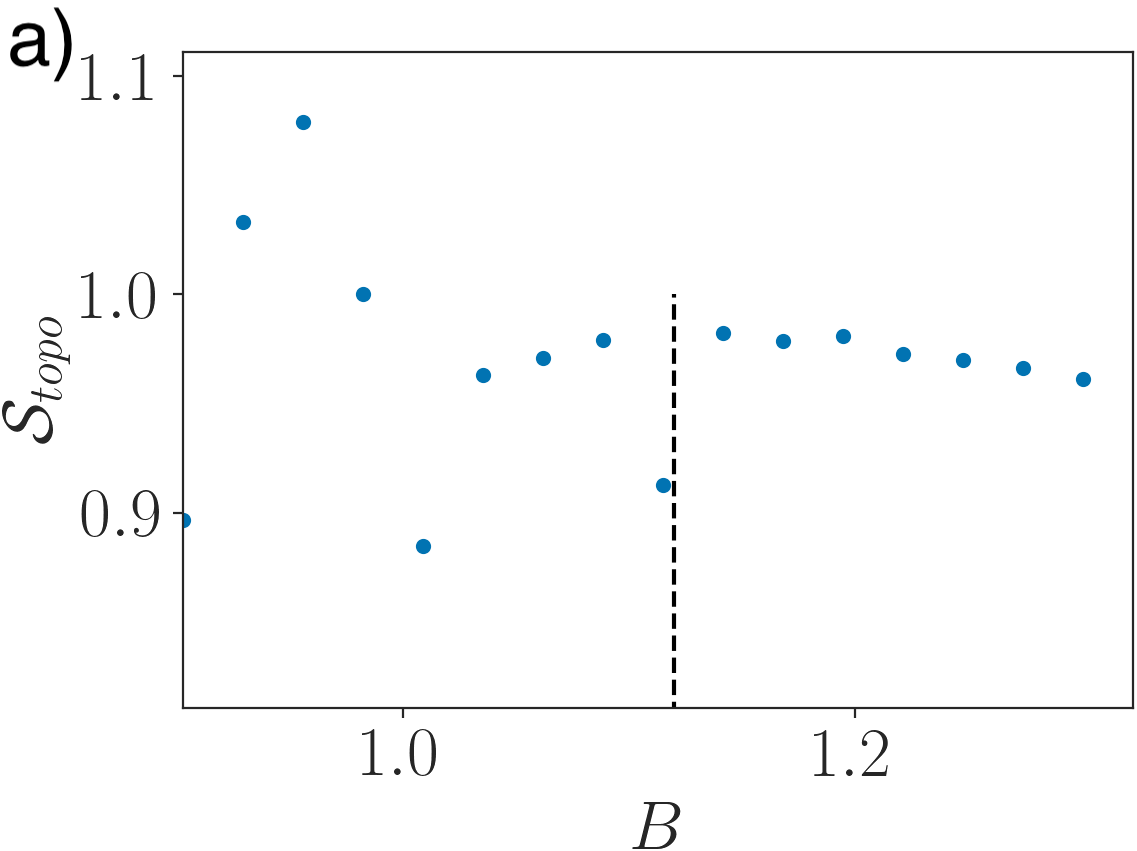} & \includegraphics[width=.48\linewidth]{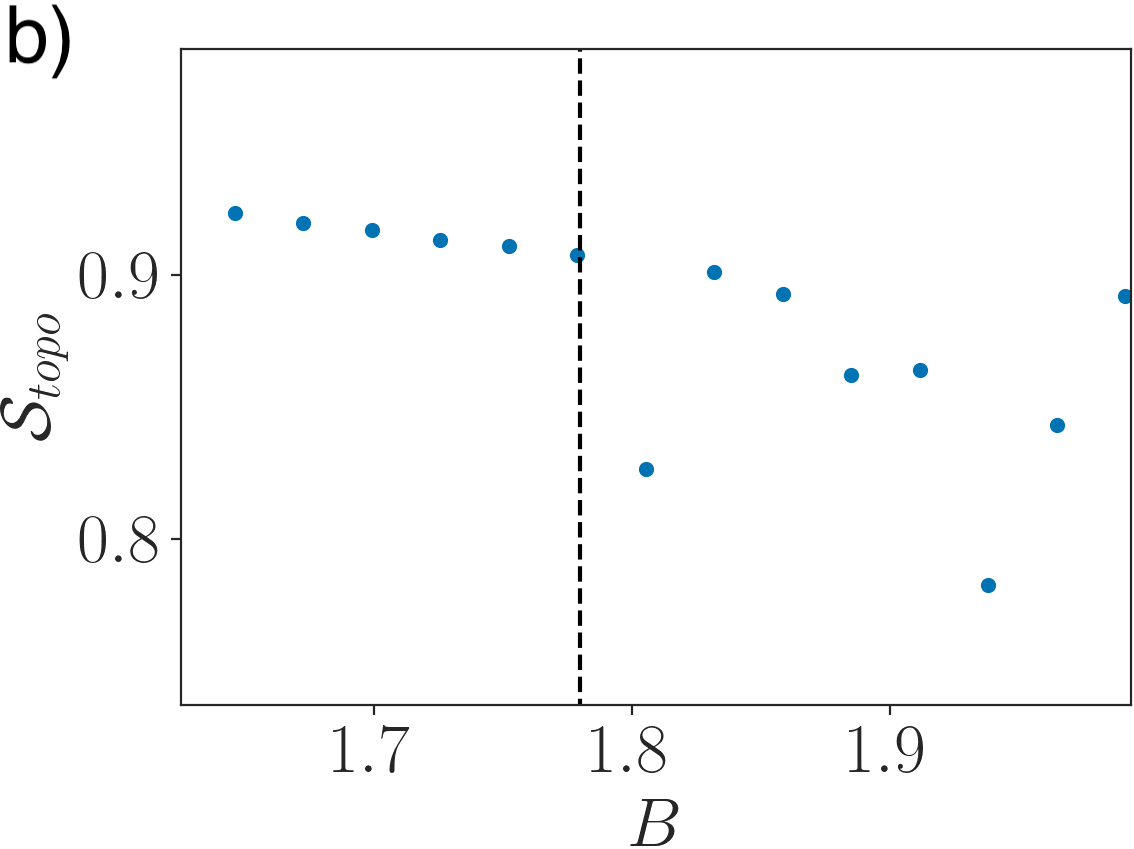} \\
        \includegraphics[width=.48\linewidth]{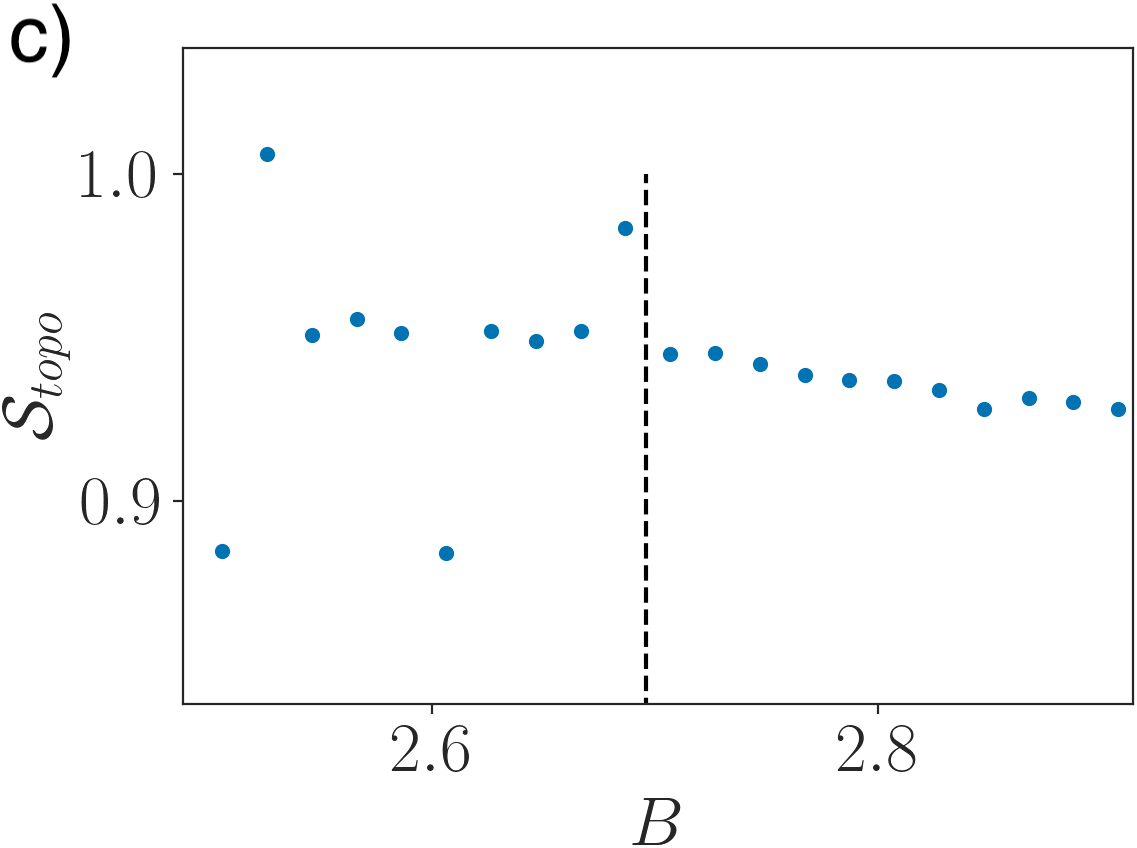} &
         \includegraphics[width=.48\linewidth]{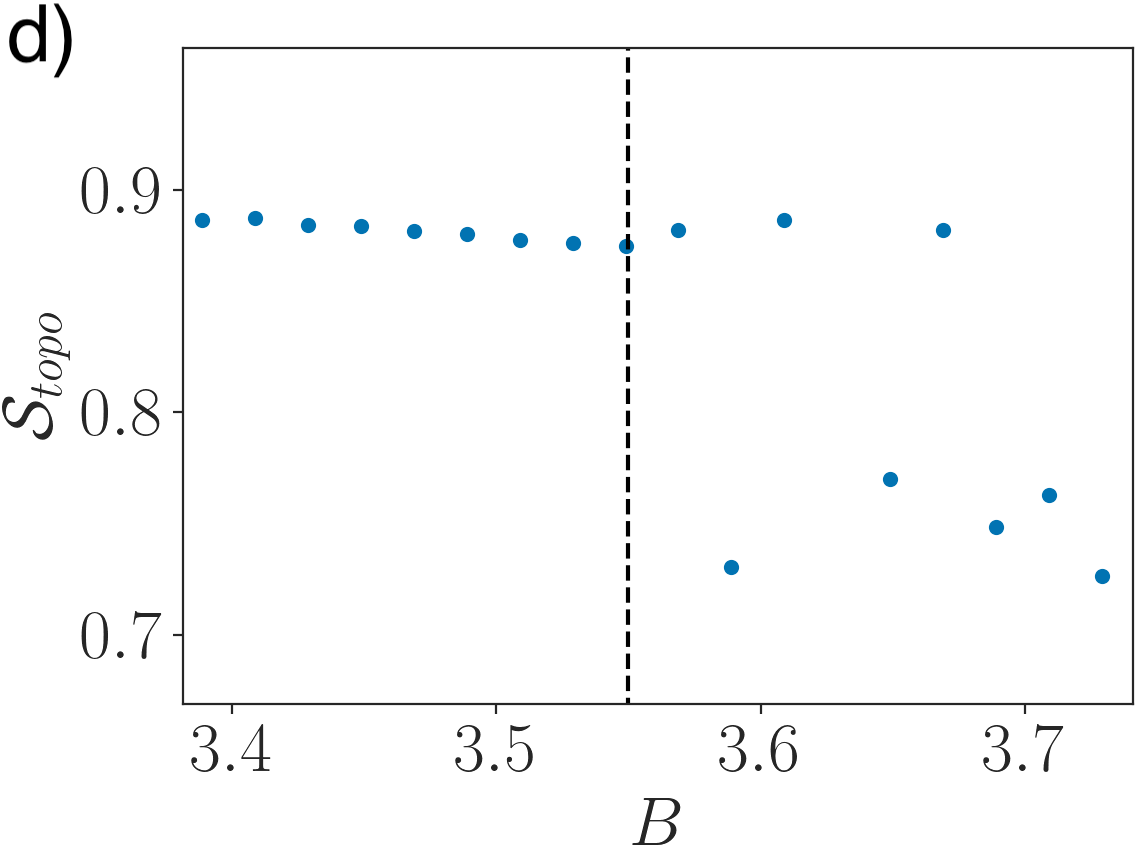}
    \end{tabular}

    \caption{(a-b) shows the left (a) and right (b) boundary separating linear part and fluctuating parts of TEE for $J_2=0.2|J_1|$, $D=2.0|J_1|$. (c-d) shows the same for $J_2=0.8|J_1|$, $D=2.0|J_1|$.}
    \label{fig:TEE_zoom_supp}
\end{figure}

Here, we have provided zoom-in to the plots of $\mathcal{S}_{topo}$ for easy observation of the critical points. In Fig.~\ref{fig:TEE_zoom_supp}, we can see linear and fluctuating parts of $\mathcal{S}_{topo}$ partitioned by vertical dashed lines. This plot, along with $C_\psi$, can then be used to clearly mark the phase transition between helical and skyrmion phases.

\section{Relation between the correlation length and parameter $J_2$}

\begin{figure}[H]
    \centering
    \includegraphics[width=1\columnwidth]{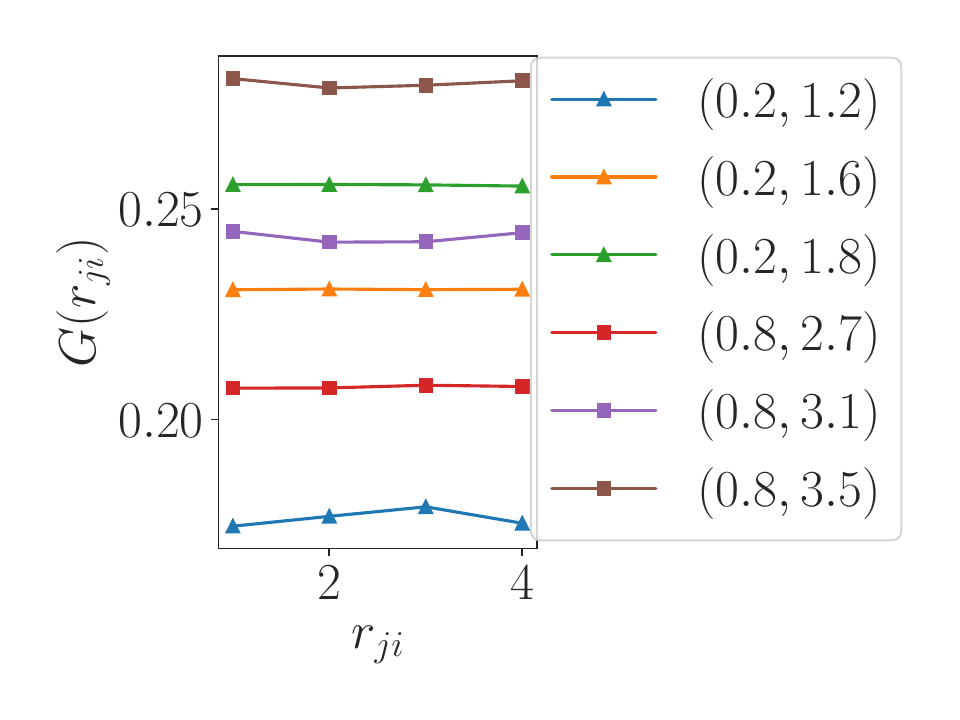}
    \caption{The correlation function $G_{||}(r_{ji})$ v/s $r_{ji}$ between sites $i$ and $j$. The legend $(J_2,B)$ represents the respective parameters in units of $|J_1|$.}
    \label{fig:correlation_function}
\end{figure}

Okubo {\it et.al.} \cite{PhysRevLett.108.017206} reported experimental and numerical results where strong further nearest neighbor interaction results in an incommensurate spiral structure, and they are stabilized under a magnetic field to give skyrmion lattice. We also identified similar properties in the theoretical model we investigated and attribute this to the system mimicking bulk when $J_2$ is large. In Fig.~\ref{fig:correlation_function} we have plotted correlation function $G(r_{ji})$ v/s $r_{ji}$ for different skyrmions formed by parameters $(J_2, B)$ in units of $|J_1|$. For larger $J_2$, the correlation function decays faster, implying a smaller correlation length. The results show that the correlation length decreases with increasing $J_2$. This indicates a lower correlation length for higher $J_2$. Shorter correlation length compared to the system boundary helps us to comply with the condition considered by Kitaev and Preskill for their definition of $S_{topo}$.

\section{Fidelity}

\begin{figure}[H]
    \centering
    \includegraphics[width=0.8\columnwidth]{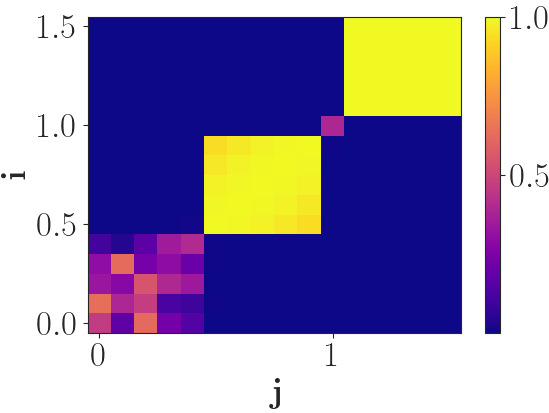}
    \caption{For $J_2=0.2|J_1|$ and $D=2.0|J_1|$ fidelity of states $<\psi_i|$ and $|\psi_j>$ corresponding to the values of magnetic field $B_i$ and $B_j$ is calculated.}
    \label{fig:fidelity}
\end{figure}

The fidelity of states corresponds to different $B$ for $J_2=0.2|J_1|$, $D=2.0|J_1|$ is shown in Fig.~\ref{fig:fidelity}. Non-zero values mostly lie along the off-diagonal, indicating the orthogonality of states in different phases.

\section{Attempt at approximations}
To eliminate finite size effects, we typically generate phase transition results with different system sizes and extrapolate to the infinite size limit. However, for the model we consider, this is computationally impractical. Therefore, we use widely used approximations. The results are shown below.

\begin{figure}
    \centering
    \includegraphics[width=1\columnwidth]{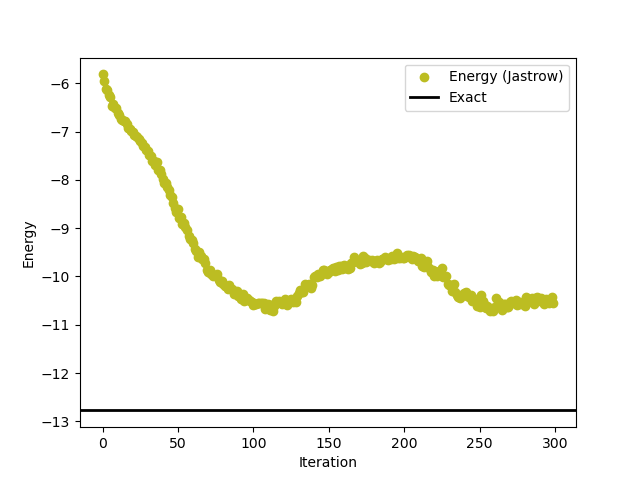}
    \caption{Montecarlo iteration v/s energy level. Black reference line is the exact solution. $D = 2.0|J_1|$, $B = 0.0|J_1|$, $J2 = 0.0|J_1|$.}
    \label{fig:montecarlo}
\end{figure}

We used the variational quantum Monte Carlo (VMC) method to approximate the ground state. The result is shown in Figure~\ref{fig:montecarlo}. VMC failed to converge for the 19-spin system we considered.
\begin{figure}
    \centering
    \includegraphics[width=1\columnwidth]{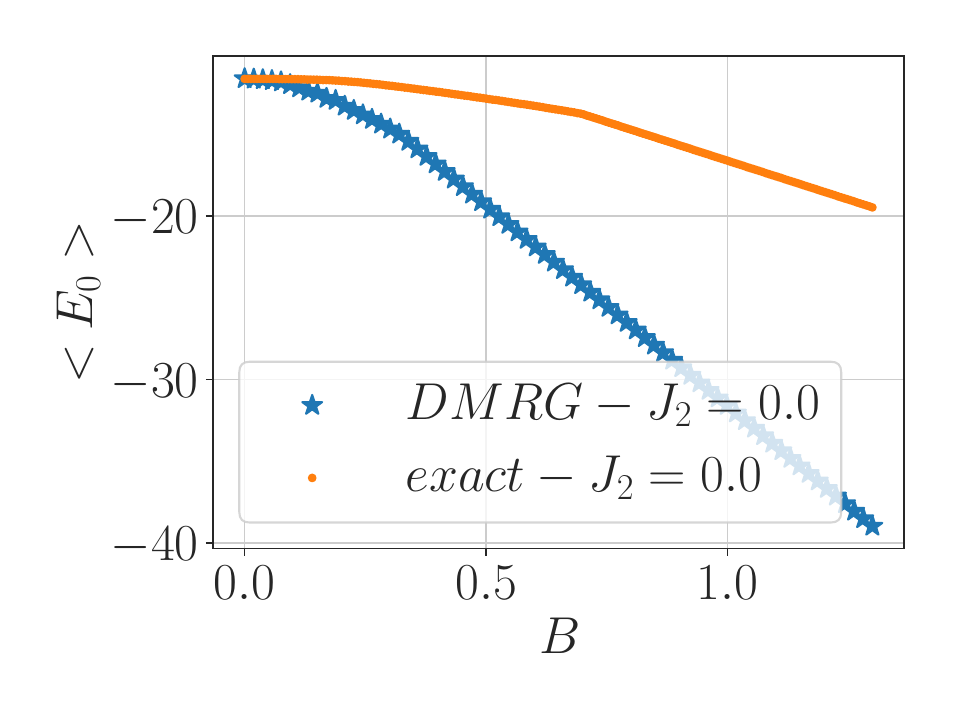}
    \caption{Comparison of the ground state energy vs. applied field using DMRG and the exact result for the 19-spin system.}
    \label{fig:DMRG}
\end{figure}

Figure~\ref{fig:DMRG} compares the ground states calculated using DMRG to the exact solution. DMRG fails to reproduce the ground state for non-zero applied field in this model.

\begin{figure}
    \centering
    \includegraphics[width=1\columnwidth]{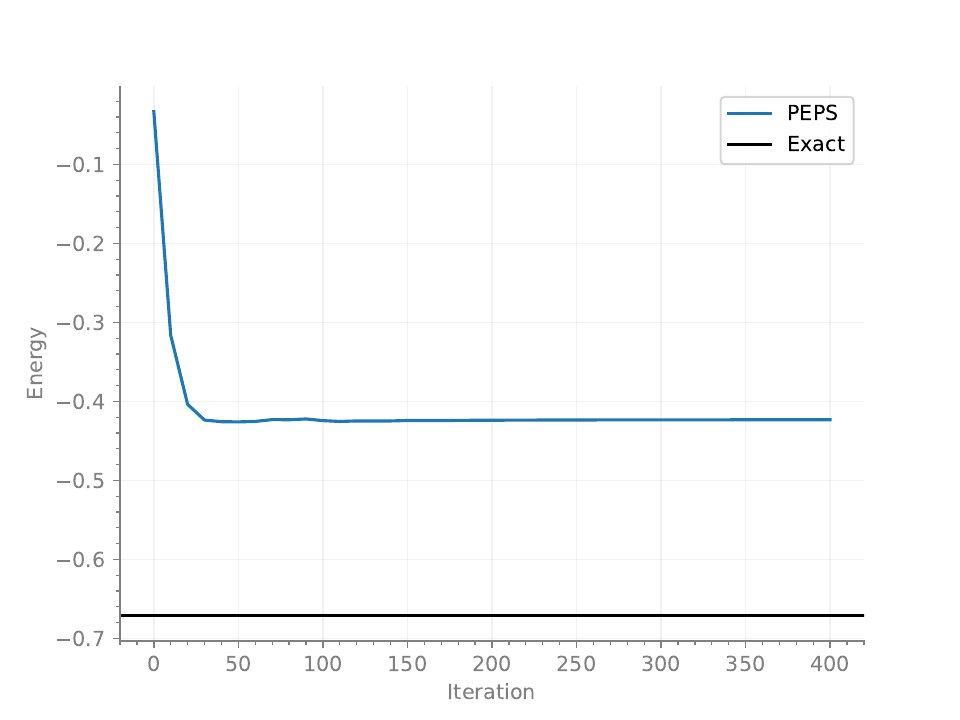}
    \caption{PEPS vs exact solution for $D = 2.0|J_1|$, $B = 0.0|J_1|$, $J2 = 0.0|J_1|$. Here we looked for iterative convergence of ground state energy per site. }
    \label{fig:PEPS}
\end{figure}
We next compared the exact solution to PEPS. As shown in Figure~\ref{fig:PEPS}, PEPS does not converge to the exact solution.

\begin{figure}
    \centering
    \includegraphics[width=1\columnwidth]{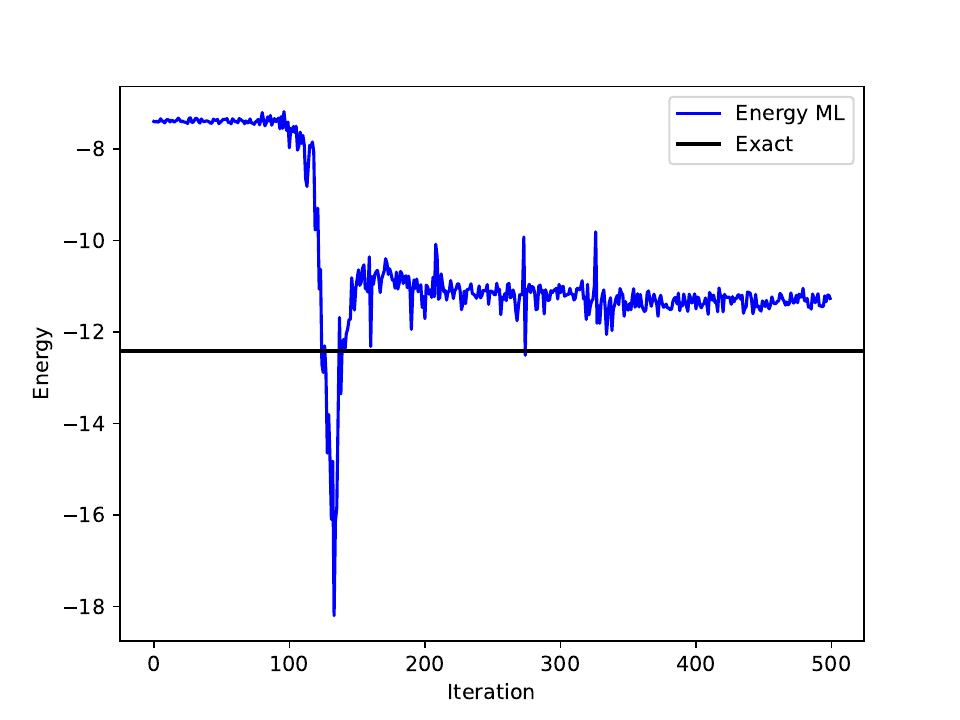}
    \caption{Neural network approximation is compared to exact solution. }
    \label{fig:netket}
\end{figure}
Finally, we used neural network quantum states to approximate the model. As shown in Figure~\ref{fig:netket}, NetKet does not converge to the exact solution for the 19-spin system. 

Because none of the approximate methods we tried converged for the 19-spin system, we cannot consider them for calculations on larger systems.